# Quantifying the Plasmonic Character of Optical Excitations in a Molecular J-Aggregate

Michele Guerrini,[†,‡] Arrigo Calzolari,[‡] Daniele Varsano,[‡] and Stefano Corni*,[‡,§]

[†]Dipartimento FIM, Università di Modena e Reggio Emilia, I-41125 Modena, Italy
[‡]CNR Nano Istituto Nanoscienze, Centro S3, I-41125 Modena, Italy
[§]Dipartimento di Scienze Chimiche, Università di Padova, Padova 35131, Italy

Ⓢ Supporting Information

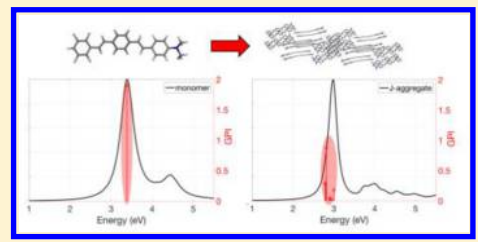

**ABSTRACT:** The definition of plasmon at the microscopic scale is far from being understood. Yet, it is very important to recognize plasmonic features in optical excitations, as they can inspire new applications and trigger new discoveries by analogy with the rich phenomenology of metal nanoparticle plasmons. Recently, the concepts of plasmonicity index and the generalized plasmonicity index (GPI) have been devised as computational tools to quantify the plasmonic nature of optical excitations. The question may arise whether any strong absorption band, possibly with some sort of collective character in its microscopic origin, shares the status of plasmon. Here we demonstrate that this is not always the case, by considering a well-known class of systems represented by J-aggregates molecular crystals, characterized by the intense J band of absorption. By means of first-principles simulations, based on a many-body perturbation theory formalism, we investigate the optical properties of a J-aggregate made of push−pull organic dyes. We show that the effect of aggregation is to lower the GPI associated with the J-band with respect to the isolated dye one, which corresponds to a nonplasmonic character of the electronic excitations. In order to rationalize our finding, we then propose a simplified one-dimensional theoretical model of the J-aggregate. A useful microscopic picture of what discriminates a collective molecular crystal excitation from a plasmon is eventually obtained.

## ■ INTRODUCTION

The identification and quantification of plasmonic excitations in nano- and molecular systems is an important issue in the field of nanoplasmonics. In fact, the rich physics and technological potential of plasmonic metal nanoparticles have triggered the question whether other nanosystems also possess excitations with such plasmonic character, to some degree, and thus a similar phenomenology. Graphene and its nanoconfined structures (polycyclic aromatic hydrocarbons in the molecular limit) attracted a great deal of attention in this respect.[1,2] This important motivation goes along with the goal of establishing the characters of localized surface plasmon excitations at a deeper microscopic level than the widespread treatment based on continuum electrodynamics. A relevant point to assess in this context is whether any strong absorption band with some sort of collective character in its microscopic origin turns out to have a plasmonic character at this scale. A positive answer (i.e., all strong excitations are recognized as plasmonic) would in fact rather belittle the quest for plasmon-like excitations, making the concept of plasmonic character of a nanoscale excitation itself redundant. Ideal systems to be investigated in this respect are J-aggregates.

J-aggregates are a class of molecular crystals with special known optical properties as an intense, narrow absorption peak (known as J-band) that appears at low energy where the single isolated monomer unit has almost zero absorption[3] and the ability to give delocalized excitons.[4] These features have been thoroughly studied in the literature[4−7] and are the result of the (strong) intermolecular coupling, which causes delocalization of the electronic excitations of the single molecule over many sites of the aggregate.[8,9] Therefore, the J-band (i) is characterized by a strong absorption cross section, (ii) requires aggregation at the nanoscale, and (iii) has a microscopic origin rooted in a collective effect. In this light, J-band seems to bear a similarity with the localized surface plasmons in metallic nanoparticles (i.e., high absorption cross section, a prominent role in nanoscopic systems and a collective origin).[10−15]

The focus of this work is specifically to quantify whether the J-band in the solid-state J-aggregate has a larger or smaller plasmonic character than its corresponding single isolated molecule. In doing this, we will respond to the underlying question whether the J-band implies a plasmon-like response. To quantify the plasmonic character, we shall use the recently proposed generalized plasmonicity index[16] (GPI). The GPI is a frequency dependent adimensional function, that has been thoroughly tested in a wide range of nanostructures, including metallic nanoparticles, silicon clusters, graphene nanostructures, and PAHs at various levels of theory (classical electrodynamics, tight-binding, jellium, and atomistic TDDFT).[16] It improves over the plasmonicity index (PI) previously proposed by Bursi et al.[17] to quantify the difference between plasmonic and single-particle electronic excitations in finite structures. In the present work not only we exploit the GPI but also we move one step further in two methodological aspects: on the one hand, we shall use an alternative level of theory to TDDFT, namely many-body perturbation theory (MBPT), which is based on the Green's function formalism and has a higher accuracy. On the other hand, we shall compute the GPI also for a macroscopic extended system such as a molecular J-aggregate,[6,18,19] besides the single molecule it is composed of.

We focus in particular on a J-aggregate molecular crystal composed of the organic conjugated dye 4-(N,N-dimethylamino)-4-(2,3,5,6-tetrafluorostyryl)stilbene[20] (see Figure 1).

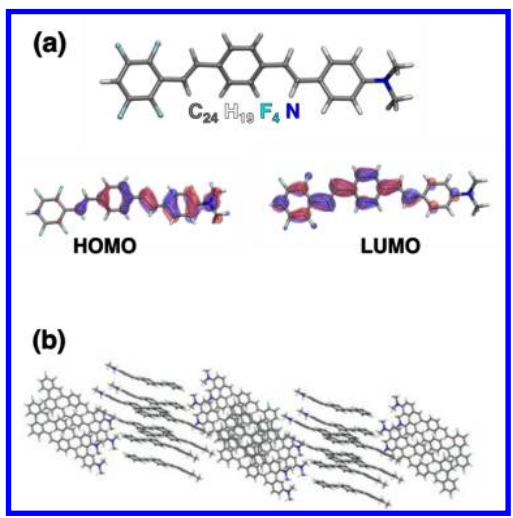

Figure 1. Atomic structure of (a) 4-(N,N-dimethylamino)-4-(2,3,5,6-tetrafluorostyryl)stilbene push–pull single molecule and (b) partial 3D view of the J-aggregate molecular crystal composed of push–pull organic dyes investigated in this work. Insets in panel (a) show isosurface plots of the highest occupied (i.e., HOMO) and lowest unoccupied (i.e., LUMO) molecular orbitals of the single molecule evaluated at the DFT level (cam-b3lyp xc-functional).

This dye is a push–pull system[21,22] that possesses an intrinsic static electric dipole due to the presence of the dimethylamino group (push) and the fluorinated aromatic ring (pull) group. We have chosen to investigate this particular push–pull organic dye for several reasons: (i) this is a realistic complex system yet not too demanding to be simulated by accurate computational techniques; (ii) the availability of its J-aggregate experimental X-ray crystal structure and of the optical absorption spectra;[20] and (iii) its charge neutrality that makes it computationally convenient.

## ■ METHOD

**Generalized Plasmonicity Index at the MBPT Level.** The GPI[16] $\eta(\omega)$ is a functional of the induced charge density $\delta n(r,\omega)$ and of the external potential $v_{ext}(r',\omega)$ for a fixed excitation frequency $\omega$. For convenience, we report here its formal expression

$$\eta(\omega) = \left| \frac{\int v_{ind}^*(r, \omega) \delta n(r, \omega) d^3 r}{\int v_{ext}^*(r', \omega) \delta n(r', \omega) d^3 r'} \right| \qquad (1)$$

where the induced Coulomb potential $v_{ind}(r,\omega)$ and the linear induced charge density $\delta n(r,\omega)$ are reported in eqs 2 and 3, respectively

$$v_{ind}(r, \omega) = \int \frac{\delta n(r, \omega)}{|r - r'|} d^3 r' \qquad (2)$$

$$\delta n(r, \omega) = \int \chi(r, r', \omega) v_{ext}(r', \omega) d^3 r' \qquad (3)$$

The charge density variation at first order in eq 3 is obtained through the density–density correlation function causal kernel[23] $\chi(r,r',\omega)$. The poles of the latter (i.e., the resonance frequencies $\omega_\xi$) give the electronic excitation energies $\hbar\omega_\xi$ of the system. When evaluated at a given resonance frequency $\omega_\xi$, the GPI can be expressed as $\eta(\omega_\xi) \approx \Gamma^{-1} E_\xi^{plas}$, where $\Gamma^{-1}$ is the excited state lifetime, and $E_\xi^{plas}$ is the plasmonic energy associated with the transition density $\rho_\xi(r)$, that represents the plasmonic contribution to the total excitation energy $\hbar\omega_\xi$:

$$E_\xi^{plas} = \int \frac{\rho_\xi^*(r) \rho_\xi(r')}{|r - r'|} d^3 r d^3 r'$$

$$= \sum_{ai} (C_{ai}^\xi)^* C_{a'i'}^\xi \int \frac{d_{ai}^*(r) d_{a'i'}(r')}{|r - r'|} d^3 r d^3 r' \qquad (4)$$

$$\rho_\xi(r) = \langle \xi | \hat{\Psi}^\dagger(r) \hat{\Psi}(r) | 0 \rangle = \sum_{ai} C_{ai}^\xi d_{ai}(r) = \sum_{ai} C_{ai}^\xi \varphi_a^*(r) \varphi_i(r) \qquad (5)$$

$$\hat{\Psi}(r) = \sum_n \varphi_n(r) \hat{c}_n, \qquad C_{ai}^\xi = \langle \xi | \hat{c}_a^\dagger \hat{c}_i | 0 \rangle \qquad (6)$$

In the second part of eq 4 we have re-expressed the transition density $\rho_\xi(r)$ as in eq 5 using an electron–hole space representation where index $a(i)$ is associated with an empty (occupied) electronic state, and $C_{ai}^\xi$ is the amplitude coefficient associated with a transition from the $i$th to the $a$th electronic state.

In metal nanoparticles, the width of a plasmonic band (and thus $\Gamma^{-1}$ in the GPI) is largely determined by the Landau damping mechanism,[24] hallmark of the mixing of the plasmon with the continuum of single-particle excitations. Band broadening in molecules is instead dominated by inhomogeneous broadening plus vibronic effects, i.e., environmental and nuclei-related (not electron-related) phenomena. Here we used a fixed $\Gamma^{-1}$ value representative of electronic effects only.

In the present work we evaluate excitation energies, transition densities, and consequently plasmonic energies starting from electronic structure simulation evaluated within the many-body perturbation theory (MBPT) framework. This assures an accurate treatment of excitations in molecular crystals, especially in the description of charge transfer effects.[25,26] Plasmonic energies are calculated as in eq 4, with transition densities coming from solution of the Bethe-Salpeter equation (BSE).[27] A detailed analysis of the BSE results, including comparison with the TDDFT level of theory and with experimental data, as well as a thorough investigation of the role of charge transfer effects and local fields within the J-band can be found in ref 28.

## ■ RESULTS

**Discussion.** Figure 2 shows the MBPT optical spectra of the single molecule and the bulk J-aggregate crystal (Figure 1). We observe the formation of the characteristic J-band that is red-shifted with respect to the isolated push−pull monomer principal peak $P$, in agreement with experimental and previous TDDFT results.[20,29] The red-shifted J-band which dominates the absorption spectrum is composed not just of the principal bright transition $J_1$ at ∼3 eV, but it gathers together many other optical transitions, with lower intensity. The inspection of the transition density (TD) of the dominant excitation $J_1$ (inset of Figure 2(b)) indicates that the TD is displaced along linear molecular chains pointing along the long axis of the crystal. Such intermolecular head-to-tail dipolar arrangement is a typical manifestation of J-aggregate behavior[19] and results in a coherent alignment of each monomer transition dipole, giving the strong total oscillator strength of the main peak at ∼3 eV. Furthermore, the constituent monomers manifest also a certain degree of intramolecular charge transfer behavior, which is compatible with their push−pull character.[22,30] By inspection of Table I, we conclude that all the analyzed excited states within the J-band (i.e., $J_1$–$J_7$) have always *lower* plasmonic energy and GPI values than those corresponding to the monomer excited state $P$ = 3.73 eV (panel a).

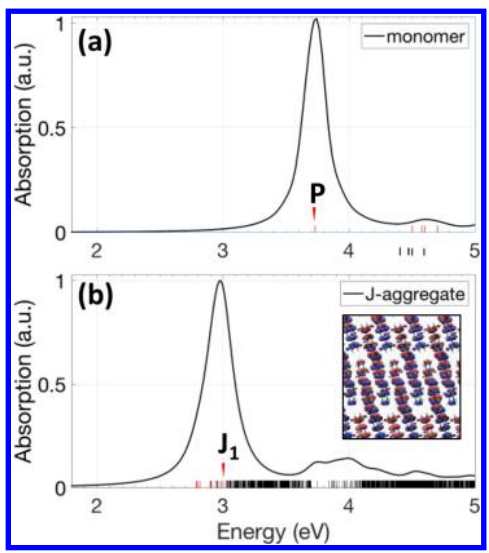

**Figure 2.** Absorption optical spectra of (a) the single monomer and (b) its molecular J-aggregate. Spectra have been normalized with respect to the maximum value. The vertical bars indicate the position of the excited states. Specific excitations analyzed in the text are marked in red. Inset of panel (b) shows the isosurface plot of the transition density associated with excited state $J_1$ belonging to the dominant peak of the J-band; red (blue) is associated with positive (negative) charge.

**Table I. Selected Excited States within J-Band, Their Associated Plasmonic Energy, Plasmonic Energy Variation $\Delta E^{plas}_{\xi_k} = E^{plas}_{\xi_k} - E^{plas}_P$, GPI Variation $\Delta\eta_{\xi_k} = \eta_{\xi_k} - \eta_P$, and Ratio $\eta_{\xi_k}/\eta_P$ with Respect to Monomer Lowest Energy Optically Active Excited State $P$ = 3.73 eV**[a]

| excited state $\xi_k$ | $E^{plas}_{\xi_k}$ | $\Delta E^{plas}_{\xi_k}$ | $\eta_{\xi_k}$ | $\Delta\eta_{\xi_k}$ | $\eta_{\xi_k}/\eta_P$ |
|---|---|---|---|---|---|
| $J_1$ | 3.000 eV | 0.072 eV | −0.608 eV | 0.18 | −1.52 | 0.1059 |
| $J_2$ | 2.951 eV | 0.008 eV | −0.672 eV | 0.02 | −1.68 | 0.0118 |
| $J_3$ | 2.905 eV | 0.020 eV | −0.660 eV | 0.05 | −1.65 | 0.0294 |
| $J_4$ | 2.897 eV | 0.011 eV | −0.833 eV | 0.03 | −1.67 | 0.0176 |
| $J_5$ | 2.813 eV | 0.349 eV | −0.331 eV | 0.87 | −0.83 | 0.5118 |
| $J_6$ | 2.796 eV | 0.111 eV | −0.570 eV | 0.28 | −1.42 | 0.1647 |
| $J_7$ | 2.786 eV | 0.116 eV | −0.560 eV | 0.29 | −1.41 | 0.1706 |

[a]Line width fixed at $\Gamma$ = 0.4 eV for all excited states; $E^{plas}_P$ = 0.68 eV and $\eta_P = \Gamma^{-1}E^{plas}_P$ = 1.7.

This is the key result of this work: the GPI signals that the collective J-band of the molecular aggregate is *less* plasmonic than the single molecule, or in other terms, the GPI identifies the J-band as nonplasmonic. This result is relevant also because it confirms the ability of the GPI to quantify the plasmonic character of excitations independently from their absorption intensity even for the molecular world (for nanoparticles it was already shown in ref 16). In other words, the GPI does not simply remap strong absorption bands into plasmonic ones but adds important independent information.

**Understanding the GPI Reduction upon Aggregation: A One-Dimensional Model Based on Molecular Exciton Theory.** We present a simple theoretical model useful to rationalize and explain the reduction in the GPI values when moving from the single isolated molecule to the molecular aggregate. For the sake of simplicity, we consider a one-dimensional aggregate (i.e., a 1D molecular chain), and we refer to its electronic excited state as *a molecular chain exciton*, i.e., an electronic excitation delocalized over several identical monomer units. Using a simple Frenkel Hamiltonian[6,19,31−34] model to describe the 1D aggregate, under periodic boundary conditions and for low lying excited states, the exciton eigenstates of an aggregate chain composed of $M$ molecules have the following form[6,31,35]

$$|\xi_k\rangle = \frac{1}{\sqrt{M}} \sum_{m=1}^{M} e^{i\frac{2\pi}{M}km}|\lambda_m\rangle, \quad -M/2 \leq k \leq M/2 \quad (7)$$

where $|\lambda_m\rangle = |\varphi^m_{exc}\rangle \prod_{n\neq m}|\varphi^n_{GS}\rangle$, $|\varphi^m_{exc}\rangle$ indicates the $m$th monomer being in an electronic excited state, and $|\varphi^n_{GS}\rangle$ indicates the $n$th monomer in the ground state. We have used eq 7, representing the molecular chain excited states, to derive the expression of the GPI within this model. In the GW/BSE simulations we assumed a fixed line width (damping) $\Gamma$. We can evaluate the molecular chain transition density (MCTD) $\rho^{agg}_k(\mathbf{r}) = \langle 0|\hat{\Psi}^\dagger(r)\hat{\Psi}(r)|\xi_k\rangle$ between the ground state $|0\rangle$ and a specific excited state $|\xi_k\rangle$ of the 1D aggregate and then use it to obtain the resulting plasmonic energy [see Section I of the Supporting Information (SI) for the derivation]

$$E^{plas}_{\xi_k} = \int \frac{\rho^{*agg}_k(r)\rho^{agg}_k(r')}{|r-r'|} dr dr' = E^{plas,mon}_{exc} + W(k) \quad (8)$$

$$W(k) = \frac{1}{M} \sum_{m\neq m'} J_{mm'} e^{i\frac{2\pi}{M}(m-m')k} \quad (9)$$

where

$$\rho^{agg}_k(r) = \frac{1}{\sqrt{M}} \sum_m e^{i\frac{2\pi}{M}km} \rho^{mon}_{exc}(r - R_m) \quad (10)$$

$$\rho^{mon}_{exc}(r - R_m) = \langle \varphi^m_{exc}|\hat{\Psi}^\dagger(r)\hat{\Psi}(r)|\varphi^m_{GS}\rangle \quad (11)$$

$$J_{mm'} = \int \frac{\rho^{*mon}_{exc}(r-R_m)\rho^{mon}_{exc}(r'-R_{m'})}{|r-r'|} dr dr' \quad (12)$$

Thus, by virtue of eq 8, the plasmonic energy associated with each excitonic level of the aggregate is given by the plasmonic energy of the monomer excited state, $E^{plas,mon}_{exc}$, plus a correction term $W(k) = \langle\xi_k|\hat{H}|\xi_k\rangle - \langle\lambda_m|\hat{H}|\lambda_m\rangle$, with $\hat{H}$ being the molecular aggregate Hamiltonian operator, whose sign and magnitude depend on the interactions between the monomers transition densities $\rho_{exc}^{mon}(r' - R_{m'})$ composing the aggregate. The energy terms in eq. 12 quantify the amount of intermolecular coupling, i.e., the energy transfer rate between $m$th and $m'$th monomers through the interaction between their transition densities.[31,36] Within the point-dipole approximation, the coupling terms $J_{mm'}$ can be expressed as [see Section I of the SI]

$$J_{mm'} = \frac{K_{mm'} - 3(\boldsymbol{\mu}_m^{mon}\cdot\bar{R})(\boldsymbol{\mu}_{m'}^{mon}\cdot\bar{R})}{R_{mm'}^3} \quad (13)$$

where $K_{mm'} = \boldsymbol{\mu}_m^{mon} \cdot \boldsymbol{\mu}_{m'}^{mon}$ and $\boldsymbol{\mu}_m^{mon} = \langle \varphi_{exc}^m | \hat{\Psi}^\dagger(r) r \hat{\Psi}(r) | \varphi_{GS}^m \rangle$ is the $m$th monomer transition dipole, $R_{mm'}$ is the distance between centers of the $m$th and $m'$th monomers, and $\bar{R}$ is the associated unit vector. Assuming all equal transition dipoles and with the same relative orientation (i.e., $\boldsymbol{\mu}_m^{mon} = \boldsymbol{\mu}$ and $K_{mm'} = K$), one can demonstrate [see Section I of the SI] that for a linear periodic chain aggregate there is a nonvanishing total transition dipole only for the excited state $\xi_0$ (i.e., $k = 0$), which in fact corresponds to the J-band of this simple 1D aggregate.

We can now proceed to evaluate the GPI variation between the two states $|\xi_0\rangle$ and $|\varphi_{exc}\rangle$, which are the lowest energy excited states of the molecular chain and the isolated monomer, respectively. For $|\xi_0\rangle$ the total transition dipole results as a coherent sum of the molecular transition dipoles, being all aligned and pointing in the same direction (the emission from this state to the ground state leads to superradiance[37]). As a rough approximation of the MCTD for $|\xi_0\rangle$, we assume a toy model composed of aligned finite-size dipoles of length $d$, arranged at a nearest neighbor distance $R = d + a$ as sketched out in the schematic visualization of Figure 3(a). Different intermolecular separations have been considered by varying the distance parameter $a$. By virtue of eq 8, the energy shift $W(k)$ physically represents the plasmonic energy variation between the monomer and the molecular chain aggregate. In turn, we can obtain the GPI variation from the single molecule excitation to the J-band as

$$\Delta \eta_{\xi_0} \approx \Gamma^{-1}(E_{\xi_0}^{plas,agg} - E_{exc}^{plas,mon}) = \Gamma^{-1}\Delta E_{\xi_0}^{plas} \quad (14)$$

$$\Delta E_{\xi_0}^{plas} = \frac{1}{M}\sum_{m\neq m'} J_{|m-m'|} = \frac{2}{M}\sum_{n=1}^{M-1}(M-n)\tilde{\tilde{J}}_n$$

$$< -\frac{4}{R}\left(\frac{M-1}{M}\right)(\gamma^2 - 1)^{-1}, \quad \gamma = \frac{R}{d} \quad (15)$$

$$J_{|m-m'|} = 2|D_{mm'}|^{-1} - |D_{mm'} + d|^{-1} - |D_{mm'} - d|^{-1} = \tilde{\tilde{J}}_{n=|m-m'|} \quad (16)$$

where $D_{mm'} = R_m - R_{m'} = (m - m')R$ is the distance between the $m$th and $m'$th monomers within the molecular aggregate. The upper limit in eq 15 comes from the nearest-neighbors approximation, i.e., the sum in eq 15 is approximated by considering only those monomer indices for which $|m - m'| = 1$.

It is straightforward to verify that the coupling terms $\tilde{\tilde{J}}_n$ in eq 16 are always negative [for further details and derivations of eqs 15 and 16 see Section II of the SI].

Figure 3(a) shows GPI variation, i.e., eq 14, as a function of the total number of monomers and considering different nearest neighbor monomer separations $R$. GPI variation is always negative, which means that the main bright transition (i.e., the J-band) in this linear J-aggregate is always *less* plasmonic than that in the isolated monomer. The reason for this behavior resides in the intrachain Coulomb couplings between monomers: when the transition dipoles of nearby molecules are parallelly aligned, the dipole–dipole interaction decreases the Coulomb energy needed to set up the overall transition density and consequently the plasmonic energy. As demonstrated by the upper bound in (15), the closer the monomers (i.e., $\gamma$ approaching 1) the more effective this effect is.

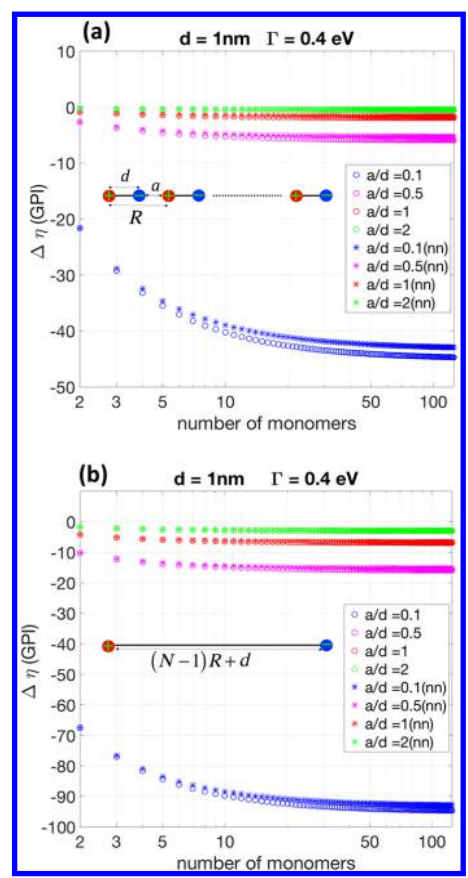

**Figure 3.** (Panel a) Difference between the GPI of a 1D J-aggregate and the GPI of the single monomer unit. (Panel b) Difference between the GPI of a 1D J-aggregate and the GPI of a quantum wire of the same length of the aggregate as sketched out in the schematic visualization. In panel a, abscissa values refer to the total number of monomers composing the molecular chain, while in panel b they indicate the length of the quantum wire, in units of intermonomer distance $R$. The colors indicate different intermolecular distances $a$. The GPI of the molecular chain is associated with the lowest excited state (i.e., $k = 0$) where all molecular transition dipoles are aligned, while the GPI of the monomer unit is associated with the lowest excited state $|\varphi_{exc}\rangle$. Plots with circles include all intersite molecular couplings, while those with asterisks have been analytically evaluated with nearest-neighbors (nn) approximation. Single monomers transition dipoles have been approximated as ideal dipoles with fixed length of $d = 1$ nm and unit charge.

To fully appreciate the difference with a clearly plasmonic transition, we compare the plasmonic energy of the lowest excited state (i.e., $k = 0$) of the 1D molecular aggregate with that of a 1D quantum wire as sketched in Figure 3(b). In order to compare the linear aggregate with the quantum wire, the two systems must have the same total oscillator strength. This have been done by making the quantum wire total transition dipole as in Figure 3(b): two point-charges of opposite signs, with the same magnitude as the ones used for each molecule in the aggregate, but placed at the ends of the wire. In Figure 3(b) we reported the difference between the GPI of the lowest transition (i.e., $\xi_0$) of the molecular aggregate and the GPI of the plasmonic quantum wire for different aggregate and wire total lengths. They are always negative, showing that indeed the J-band is far from being a plasmon. The picture that emerges from this comparison is that two excitations with similar absorption spectrum (here the J-band and the plasmonic quantum wire ones) may have a different plasmonic character: even if the positive and negative molecular induced charges, internal to the medium, cancel out in determining the overall oscillator strength of the J-band, they do not for plasmonic energy and the GPI.

## CONCLUSIONS

In this work we have evaluated the generalized plasmonicity index of a push−pull organic dye and of its bulk J-aggregate molecular crystal, by using a many-body perturbation theory approach. Our results indicate that the electronic excitations within the characteristic red-shifted J-band have a lower plasmonic character with respect to that of the single monomer. Despite the fact that the J-band is a collective excitation, it does not have a plasmonic character. In order to explain and rationalize the MBPT result, we proposed a simple one-dimensional model to evaluate analytically the plasmonic energy of the extended molecular system. The proposed model is able to predict the lowering of the plasmonic energy as due to a globally negative interaction energy between monomers transition densities along the chain (i.e., intrachain couplings), which is also the origin of the characteristic J-band red-shift.

In a more general perspective, we have clearly shown that a bright excitation in molecular systems, even when coming from a collective mechanism, may well have no relation with plasmons. This highlights the special status of molecular excitations that do present plasmon-like character.[1,2,16,38−43]

## COMPUTATIONAL DETAILS

The unit cell structure of the bulk molecular J-aggregate has been taken from available X-ray diffraction measurements[20] (available in the CCDC No. 961738). To simulate the isolated (i.e., in gas-phase) push−pull dye, we extracted one monomer unit from the bulk J-aggregate crystal and then relaxed its structure with Gaussian[44] using the cam-b3lyp xc-functional and fixing a force threshold of $10^{-5}$ Hartree/Bohr. Then, with the MOLGW[45] code we performed a DFT ground state calculation using the cam-b3lyp xc-functional and the cc-pVDZ localized basis set with frozen core approximation, then we applied the quasi-particle corrections to the DFT energy levels by using an eigenvalue-only self-consistent GW[45] procedure, and, at the end, we evaluated the BSE optical absorption spectrum using the Tamm-Dancoff approximation.

To simulate the bulk molecular crystal, electronic and optical properties of the J-aggregate have been obtained by using the Quantum Espresso[46] (QE) package in conjunction with the plane wave MBPT code YAMBO.[47] Using PBE[48] xc-functional and norm-conserving pseudopotentials,[49] with QE we calculated the ground state DFT electronic structure of the aggregate. Then, with YAMBO we applied the GW quasi-particle correction to the DFT ground state energies using the plasmon pole approximation.[50] At the end, the J-aggregate optical absorption spectrum has been obtained through solution of the Bethe-Salpeter equation[27,51,52] (BSE) in the Tamm-Dancoff approximation. Through direct diagonalization of the BSE Hamiltonian we have extracted, for each excited state $|\xi_k\rangle$, the transition coefficients $C_{ai}^{\xi_k}$ that enter in the expression of the transition density (eq 5) for each excited state and consequently of the plasmonic energy (eq 4) and the GPI at resonance. In order to evaluate the plasmonic energy for each excited state, a specific postprocessing routine has been implemented both in MOLGW and in YAMBO.

Due to the strong anisotropy of the molecular bulk aggregate, the final optical absorption spectrum reported in Figure 2(b) has been evaluated by performing three independent simulations with three orthogonal external exciting field orientations and then taking the spatial average over the three obtained spectra, i.e., $\frac{1}{3}\cdot\mathfrak{I}Tr[\varepsilon_{M,ij}(\omega)]$, where $\mathfrak{I}$ stands for the *imaginary part of*, and $\varepsilon_{M,ij}(\omega)$ is the macroscopic, frequency dependent, dielectric tensor in the long wavelength limit.

## ASSOCIATED CONTENT

### Supporting Information

The Supporting Information is available free of charge on the ACS Publications website at DOI: 10.1021/acs.jctc.9b00220.

> Section I, expression of polymer transition density together with point-dipole approximation of intermolecular energy transfer integral, i.e., eqs 10 and 12 (main text), respectively; Section II, GPI variation from single molecule to one-dimensional J-aggregate model, i.e., eqs 14−16 (main text) (PDF)

## AUTHOR INFORMATION

### Corresponding Author
*E-mail: stefano.corni@unipd.it.
### ORCID
Arrigo Calzolari: 0000-0002-0244-7717
Daniele Varsano: 0000-0001-7675-7374
Stefano Corni: 0000-0001-6707-108X
### Funding
This work was partially funded by the European Union under the ERC grant TAME Plasmons (ERC-CoG- 681285).
### Notes
The authors declare no competing financial interest.

## ACKNOWLEDGMENTS

The authors would like to thank Dr. Luca Bursi of the Physics and Astronomy Department at Rice University, Houston, TX, for useful discussions and the critical reading of the manuscript and Prof. Peter Nordlander of the Physics and Astronomy Department at Rice University, Houston, TX, for useful discussions.

## REFERENCES

(1) Chapkin, K. D.; Bursi, L.; Stec, G. J.; Lauchner, A.; Hogan, N. J.; Cui, Y.; Nordlander, P.; Halas, N. J. Lifetime Dynamics of Plasmons in the Few-Atom Limit. *Proc. Natl. Acad. Sci. U. S. A.* **2018**, *115*, 9134−9139.
(2) Lauchner, A.; Schlather, A. E.; Manjavacas, A.; Cui, Y.; McClain, M. J.; Stec, G. J.; García De Abajo, F. J.; Nordlander, P.; Halas, N. J. Molecular Plasmonics. *Nano Lett.* **2015**, *15*, 6208−6214.

(3) Eisfeld, A.; Briggs, J. S. The J-Band of Organic Dyes: Lineshape and Coherence Length. *Chem. Phys.* **2002**, *281*, 61−70.
(4) Würthner, F.; Kaiser, T. E.; Saha-Möller, C. R. J-Aggregates: From Serendipitous Discovery to Supramolecular Engineering of Functional Dye Materials. *Angew. Chem., Int. Ed.* **2011**, *50*, 3376−3410.
(5) Walczak, P. B.; Eisfeld, A.; Briggs, J. S. Exchange Narrowing of the J Band of Molecular Dye Aggregates. *J. Chem. Phys.* **2008**, *128*, 044505.
(6) Egorov, V. V. Theory of the J-Band: From the Frenkel Exciton to Charge Transfer. *Phys. Procedia* **2009**, *2*, 223−326.
(7) Knapp, E. W. Lineshapes of Molecular Aggregates, Exchange Narrowing and Intersite Correlation. *Chem. Phys.* **1984**, *85*, 73−82.
(8) Van Burgel, M.; Wiersma, D. A.; Duppen, K. The Dynamics of One-Dimensional Excitons in Liquids. *J. Chem. Phys.* **1995**, *102*, 20−33.
(9) Tempelaar, R.; Stradomska, A.; Knoester, J.; Spano, F. C. Anatomy of an Exciton: Vibrational Distortion and Exciton Coherence in H- and J-Aggregates. *J. Phys. Chem. B* **2013**, *117*, 457−466.
(10) Tame, M. S.; McEnery, K. R.; Özdemir, Ş. K.; Lee, J.; Maier, S. A.; Kim, M. S. Quantum Plasmonics Review. *Nat. Phys.* **2013**, *9*, 329−340.
(11) Stiles, P. L.; Dieringer, J. A.; Shah, N. C.; Van Duyne, R. P. Surface-Enhanced Raman Spectroscopy. *Annu. Rev. Anal. Chem.* **2008**, *1*, 601−626.
(12) Archambault, A.; Marquier, F.; Greffet, J. J.; Arnold, C. Quantum Theory of Spontaneous and Stimulated Emission of Surface Plasmons. *Phys. Rev. B: Condens. Matter Mater. Phys.* **2010**, *82*, 035411.
(13) Pitarke, J. M.; Silkin, V. M.; Chulkov, E. V.; Echenique, P. M. *Rep. Prog. Phys.* **2007**, *70*, 1−87.
(14) Moradi, A. Surface and Bulk Plasmons of Electron-Hole Plasma in Semiconductor Nanowires. *Phys. Plasmas* **2016**, *23*, 114503.
(15) Novotny, L.; Hecht, B. Surface Plasmons. In *Principles of Nano-Optics*, 1st ed.; Cambridge University Press: 2006; pp 393−410,.
(16) Zhang, R.; Bursi, L.; Cox, J. D.; Cui, Y.; Krauter, C. M.; Alabastri, A.; Manjavacas, A.; Calzolari, A.; Corni, S.; Molinari, E.; et al. How to Identify Plasmons from the Optical Response of Nanostructures. *ACS Nano* **2017**, *11*, 7321−7335.
(17) Bursi, L.; Calzolari, A.; Corni, S.; Molinari, E. Quantifying the Plasmonic Character of Optical Excitations in Nanostructures. *ACS Photonics* **2016**, *3*, 520−525.
(18) Spano, F. C.; Silva, C. H- and J-Aggregate Behavior in Polymeric Semiconductors. *Annu. Rev. Phys. Chem.* **2014**, *65*, 477−500.
(19) Eisfeld, A.; Briggs, J. S. The J- and H-Bands of Organic Dye Aggregates. *Chem. Phys.* **2006**, *324*, 376−384.
(20) Botta, C.; Cariati, E.; Cavallo, G.; Dichiarante, V.; Forni, A.; Metrangolo, P.; Pilati, T.; Resnati, G.; Righetto, S.; Terraneo, G.; et al. Fluorine-Induced J-Aggregation Enhances Emissive Properties of a New NLO Push-Pull Chromophore. *J. Mater. Chem. C* **2014**, *2*, 5275−5279.
(21) Baumeier, B.; Rohlfing, M.; Andrienko, D. Electronic Excitations in Push-Pull Oligomers and Their Complexes with Fullerene from Many-Body Green's Functions Theory with Polarizable Embedding. *J. Chem. Theory Comput.* **2014**, *10*, 3104−3110.
(22) Panja, S. K.; Dwivedi, N.; Saha, S. Tuning the Intramolecular Charge Transfer (ICT) Process in Push−Pull Systems: Effect of Nitro Groups. *RSC Adv.* **2016**, *6*, 105786−105794.
(23) Fetter, A. L.; Walecka, J. D. *Quantum Theory of Many-Particle Systems*, 1st ed.; McGraw Hill Book Company: 1971.
(24) Li, X.; Xiao, D.; Zhang, Z. Landau Damping of Quantum Plasmons in Metal Nanostructures. *New J. Phys.* **2013**, *15*, 023011.
(25) Baumeier, B.; Andrienko, D.; Rohlfing, M. Frenkel and Charge-Transfer Excitations in Donor−Acceptor Complexes from Many-Body Green's Functions Theory. *J. Chem. Theory Comput.* **2012**, *8*, 2790−2795.
(26) Blase, X.; Attaccalite, C. Charge-Transfer Excitations in Molecular Donor-Acceptor Complexes within the Many-Body Bethe-Salpeter Approach. *Appl. Phys. Lett.* **2011**, *99*, 171909.
(27) Onida, G.; Reining, L.; Rubio, A. Electronic Excitations: Density-Functional versus Many-Body Green's-Function Approaches. *Rev. Mod. Phys.* **2002**, *74*, 601−659.
(28) Guerrini, M.; Cocchi, C.; Calzolari, A.; Varsano, D.; Corni, S. Interplay between Intra- and Inter-Molecular Charge Transfer in the Optical Excitations of J-Aggregates. *J. Phys. Chem. C* **2019**, *123*, 6831−6838.
(29) Guerrini, M.; Calzolari, A.; Corni, S. Solid-State Effects on the Optical Excitation of Push−Pull Molecular J-Aggregates by First-Principles Simulations. *ACS Omega* **2018**, *3*, 10481−10486.
(30) Zhu, HZ.; Xian, W.; Renjun, M.; Zhuoran, K.; Qianjin, G.; Andong, X. Intramolecular Charge Transfer and Solvation of Photoactive Molecules with Conjugated Push−Pull Structures. *ChemPhysChem* **2016**, *17*, 3245−3251.
(31) Davydov, A. S. The Theory of Molecular Excitons (Translation). *Usp. Fiz. Nauk* **1964**, *82*, 393−448.
(32) Barford, W.; Marcus, M. Theory of Optical Transitions in Conjugated Polymers. I. Ideal Systems. *J. Chem. Phys.* **2014**, *141*, 164101.
(33) Marcus, M.; Tozer, O. R.; Barford, W. Theory of Optical Transitions in Conjugated Polymers. II. Real Systems. *J. Chem. Phys.* **2014**, *141*, 164102.
(34) Eisfeld, A.; Briggs, J. S. The J-Band of Organic Dyes: Lineshape and Coherence Length. *Chem. Phys.* **2002**, *281*, 61−70.
(35) Kasha, M.; Rawls, H. R.; Ashraf El-Bayoumi, M. The Exciton Model in Molecular Spectroscopy. *Pure Appl. Chem.* **1965**, *11*, 371−392.
(36) Kasha, M. Energy Transfer Mechanisms and the Molecular Exciton Model for Molecular Aggregates. *Radiat. Res.* **1963**, *20*, 55−70.
(37) Spano, F. C.; Mukamel, S. Superradiance in Molecular Aggregates. *J. Chem. Phys.* **1989**, *91*, 683−700.
(38) Manjavacas, A.; Marchesin, F.; Thongrattanasiri, S.; Koval, P.; Nordlander, P.; Sánchez-Portal, D.; García De Abajo, F. J. Tunable Molecular Plasmons in Polycyclic Aromatic Hydrocarbons. *ACS Nano* **2013**, *7*, 3635−3643.
(39) Krauter, C. M.; Schirmer, J.; Jacob, C. R.; Pernpointner, M.; Dreuw, A. Plasmons in Molecules: Microscopic Characterization Based on Orbital Transitions and Momentum Conservation. *J. Chem. Phys.* **2014**, *141*, 104101.
(40) Bernadotte, S.; Evers, F.; Jacob, C. R. Plasmons in Molecules. *J. Phys. Chem. C* **2013**, *117*, 1863−1878.
(41) Guidez, E. B.; Aikens, C. M. Quantum Mechanical Origin of the Plasmon: From Molecular Systems to Nanoparticles. *Nanoscale* **2014**, *6*, 11512−11527.
(42) Bursi, L.; Calzolari, A.; Corni, S.; Molinari, E. Light-Induced Field Enhancement in Nanoscale Systems from First-Principles: The Case of Polyacenes. *ACS Photonics* **2014**, *1*, 1049−1058.
(43) Guidez, E. B.; Aikens, C. M. Origin and TDDFT Benchmarking of the Plasmon Resonance in Acenes. *J. Phys. Chem. C* **2013**, *117*, 21466−21475.
(44) Frisch, M. J.; Trucks, G. W.; Schlegel, H. B.; Scuseria, G. E.; Robb, M. A.; Cheeseman, J. R.; Montgomery, J. A.; Vreven, T.; Kudin, K. N.; Burant, J. C.; et al. *Gaussian 09*, Revision A.02; 2016.
(45) Bruneval, F.; Rangel, T.; Hamed, S. M.; Shao, M.; Yang, C.; Neaton, J. B. Molgw 1: Many-Body Perturbation Theory Software for Atoms, Molecules, and Clusters. *Comput. Phys. Commun.* **2016**, *208*, 149−161.
(46) Giannozzi, P.; Baroni, S.; Bonini, N.; Calandra, M.; Car, R.; Cavazzoni, C.; Ceresoli, D.; Chiarotti, G. L.; Cococcioni, M.; Dabo, I.; et al. QUANTUM ESPRESSO: A Modular and Open-Source Software Project for Quantum Simulations of Materials. *J. Phys.: Condens. Matter* **2009**, *21*, 395502.
(47) Marini, A.; Hogan, C.; Grüning, M.; Varsano, D. Yambo: An Ab Initio Tool for Excited State Calculations. *Comput. Phys. Commun.* **2009**, *180*, 1392−1403.

(48) Perdew, J. P.; Burke, K.; Ernzerhof, M. Generalized Gradient Approximation Made Simple. *Phys. Rev. Lett.* **1996**, *77*, 3865−3868.

(49) Hamann, D. R.; Schlüter, M.; Chiang, C. Norm-Conserving Pseudopotentials. *Phys. Rev. Lett.* **1979**, *43*, 1494−1497.

(50) Larson, P.; Dvorak, M.; Wu, Z. Role of the Plasmon-Pole Model in the GW Approximation. *Phys. Rev. B: Condens. Matter Mater. Phys.* **2013**, *88*, 125205.

(51) Cudazzo, P.; Gatti, M.; Rubio, A. Excitons in Molecular Crystals from First-Principles Many-Body Perturbation Theory: Picene versus Pentacene. *Phys. Rev. B: Condens. Matter Mater. Phys.* **2012**, *86*, 195307.

(52) Marques, M. A. L.; Ullrich, C. A.; Nogueira, F.; Rubio, A.; Burke, K.; Gross, E. K. U. *Time-Dependent Density Functional Theory*, 1st ed.; Springer: 2006;.